\documentstyle[12pt,aasms4,epsf]{article}
\def\ee #1 {\times 10^{#1}}          % \ee p       10^p
\def\ut #1 #2 { \, \textrm{#1}^{#2}} % \ut unit p  unit^p
\def\u #1 { \, \textrm{#1}}          % \u unit     unit

\def\dsec   {\hbox{$.\!\!^{\rm s}$}}            % Second over dot
\def\lsol{\, \hbox{$\hbox{L}_\odot$}}

\begin{document}
\title{Nonthermal Emission from the Arches Cluster (G0.121+0.017) and
the Origin of $\gamma$-ray Emission from 3EG J1746-2851}

\author{F. Yusef-Zadeh\footnote{Department of Physics and Astronomy,
Northwestern University,
Evanston, Il. 60208 (zadeh@northwestern.edu)},
 M. Nord\footnote{Naval Research Lab., Washington, DC 20375-5351
(Michael.Nord@nrl.navy.mil)}, M. Wardle\footnote {Department of Physics,
Macquarie University, NSW 2109, Australia
(wardle@physics.mq.edu.au)}, 
C. Law\footnote{Department of Physics and Astronomy, Northwestern
University,
Evanston, Il. 60208 (claw@northwestern.edu)}, C. Lang
\footnote{Department of Physics \& Astronomy, University
of Iowa, Iowa City, IA 52245
(cornelia-lang@uiowa.edu)}, 
T.J.W. Lazio\footnote{Naval Research Lab., Washington, DC 20375-5351
(Joseph.Lazio@nrl.navy.mil)}}

\begin{abstract}

High resolution VLA observations of the Arches cluster near the
Galactic center show evidence of continuum emission at $\lambda$3.6,
6, 20 and 90cm.  The continuum emission at $\lambda$90cm is
particularly striking because thermal sources generally become
optically thick at longer wavelengths and fall off in brightness
whereas non-thermal sources increase in brightness.  It is argued that
the radio emission from this unique source has compact and diffuse
components produced by thermal and nonthermal processes, respectively.
Compact sources within the cluster arise from stellar winds of
mass-losing stars (Lang, Goss \& Rodriguez 2001a) whereas diffuse
emission is likely to be due to colliding wind shocks of the cluster
flow generating relativistic particles due to diffuse shock
acceleration.  We also discuss the possibility that $\gamma$-ray
emission from 3EG J1746--2851, located within 3.3$'$ of the Arches
cluster, results from the inverse Compton scattering of the radiation
field of the cluster.

\end{abstract}

\keywords{ISM: abundances---ISM: cosmic-rays---The Galaxy: center---Xrays: ISM}

%\vfill\eject

\section{Introduction}

Near-IR observations of the Galactic center region have identified a
remarkable young (1--2 Myr) cluster of massive stars, the Arches
cluster or G0.121+0.017 (e.g., Cotera et al.  1996;
Figer et al.  2002).  The Arches cluster has 
an angular size of $\sim15''$ with the greatest concentration of stars
lying in the inner 9$''$ (0.36 pc at the distance of 8.5 kpc).  
Recently, radio continuum emission from individual stars in the Arches
cluster was detected between 45 and 8.5 GHz (Lang et al.
2001a).  The radio spectra, near-IR spectral types of several stars
in the Arches cluster  and X-ray studies of this cluster are all 
consistent with colliding ionized stellar winds
arising from mass-losing WN and/or Of stars (Cant\'o et al.\
2000; Raga et al.\ 2001; Yusef-Zadeh et al.\ 2002a).  
Here, we present sensitive
observations at 327 MHz and present evidence for nonthermal emission
from the Arches cluster.  

\section {Observations \& Results}
%\subsection{Radio data}

{\bf {Radio Data:}} The Very Large Array (VLA) of the National Radio Astronomy
Observatory\footnote{The National Radio Astronomy Observatory is a
facility of the National Science Foundation, operated under a
cooperative agreement by Associated Universities, Inc.} was used in a
number of configurations to observe the Galactic center at 330 MHz
with a spatial resolution of 7$''\times12''$ and rms noise of
$\approx$1.7 mJy.  A detailed account is given
in Nord et al.  (2003).  We also used archival 1.4, 5, 8.5 GHz 
observations of the cluster which  were
based on multiple VLA configurations (Morris \& Yusef-Zadeh 1989;
 Lang et al.\ 2001a).
Figure 1a shows a grayscale image of the Arches cluster at
327 MHz  with a resolution of 7$''\times12''$.  A bright radio
source with a flux density of 91 mJy appears to coincide with the Arches
cluster at $\alpha$,
$\delta $[2000] = $17^{\rm h} 45^{\rm m}$ 50\dsec81$\pm0.53$,
$-28^\circ 49^\prime 20.1\pm0.82^{\prime\prime}$.  Statistical
astrometry  was used to determine the positions of the point
sources in the image.  The source is resolved to a size of
7.1$''(\alpha)\times13.5''(\delta)$.  Faint filamentary structures are
detected to the
north and south of the Arches cluster (Nord et al.  2003).  Figure 1b
shows the  identical region  at
1.4 GHz  with a resolution of 3.9$''\times3.2''$.  Although the
1.4 GHz continuum emission is dominated by bright HII regions
associated with arched filaments and nonthermal filaments of the Arc,
there is localized emission from the position of the Arches cluster at
a number of frequencies including the 1.4 GHz data.  The peak
brightness of the cluster at 1.4 GHz is located at $\alpha$, $\delta
$[2000] = $17^{\rm h} 45^{\rm m}$ 50\dsec36$\pm0.09$, $-28^\circ
49^\prime 20.53\pm1.3^{\prime\prime}$, coincident with the
327 MHz  peak brightness within positional uncertainties.

{\bf {X-ray Data}} 
X-ray emission from the Arches cluster has recently been
detected by \emph{Chandra} (ObsID 945) (Yusef-Zadeh et
al. 2002a).  
Five X-ray sources (A1 to A5) were detected in the region
surrounding 
the Arches cluster with A1 being the brightest source arising 
from the core of the cluster. 
The spectral and the spatial distributions  of the A1 component of the
X-ray
emission were extracted from all available
archival 
\emph{Chandra} observations (ObsIDs 945, 2276, 2284).
The 327 MHz emission
coincides with the brightest X-ray component A1 which arises from the
core of the Arches cluster.  Figure 2 shows contours of X-ray
emission superimposed on a grayscale continuum image at 327 MHz.  It
is clear that source A1, which is now spatially resolved into two
components (Law and Yusef-Zadeh 2003), coincides with the peak of the
327 MHz emission from the Arches cluster (Yusef-Zadeh et al.  2002a).

{\bf {Radio Spectrum:}} 
The detection of the continuum emission from the Arches cluster
from 327 MHz to 43 GHz (see Fig. 3)  suggests  both
thermal and nonthermal emission.  
Here we argue that high frequency
radio emission from the cluster is compact and arises from stellar
sources, whereas the low-frequency radio emission appears diffuse with
nonthermal characteristics.

First, the prominent thermal components seen at
1.4 GHz are absent at 327 MHz  whereas the
nonthermal filaments and the Arched cluster are evident 
at both frequencies. The typical brightness of HII regions
associated with the Arched filaments drops by a factor of $\sim$20
from 1.4GHz to 327 MHz whereas the continuum emission from the Arches
cluster increases at 327 MHz.  
These suggest that the bright compact source at 327 MHz
coincident with the Arches cluster, like the linear features, has
nonthermal characteristics. 
A faint nonthermal stellar source with a flux density 0.4 mJy at 5 GHz
(AR6 in Lang
et al. 2001a) 
 has been 
detected toward the Arches cluster but its 327 MHz emission should be
insignificant. 
The thermal emission 
from ionized stellar wind 
sources should also be insignificant at low frequencies. 

Second, if the continuum source
coincident with the Arches cluster due to free-free emission
at a temperature of 10$^{4}$K, the emission measure E (cm$^{-6}$ pc)
estimated from the 8.5 GHz data is $\approx2\times10^5$
cm$^{-6}$ pc.  The 8.5 GHz  flux density is 
14.8 mJy with a resolved source size of
5.5$''\times8.7''$.  The predicted optical depth at 327 MHz is then 
$\approx$0.65. However, the observed  flux density at 327 MHz 
is much higher, indicating the presence of  nonthermal diffuse 
emission at low frequencies.  The large
free-free optical depth at 327 MHz detected toward the Arched
filaments (Anantharamaiah et al.\ 1991), but not toward the Arches
cluster and the nonthermal filaments, implies that the Arches cluster
and nonthermal filaments are on the near side of the HII region.  A
similar geometry has been suggested by Lang et al.  (2001b) based on
H92$\alpha$ observations of the Arched filaments.

The flux density of the 327 MHz emission, 
91$\pm4.8$ mJy (see Table 1),  is a reasonable estimate
of nonthermal emission from the Arches cluster because of:
 i) the low frequency, ii) properly subtracted
background emission and iii) its isolation from the rest of thermal
gas which is optically thick at 327 MHz.  The best estimate of the
thermal component is estimated from 8 GHz and higher frequency
observations.  Sub-arcsecond resolution images of the Arches cluster
show a total flux of 3.2 mJy from a collection of ionized stellar wind
sources detected at 8.5 GHz (Lang et al.\ 2001a).  A Gaussian fit with
a linear background subtraction to the 8 GHz radio continuum data
having a resolution of 6.6$''\times7.8''$ gives a flux density of
14.8$\pm$4.8 mJy (see Table 1).  This value is much higher than the
total flux density of compact stellar sources, thus implying a new
diffuse component associated with the Arches cluster.

The flux measurements of the Arches cluster at 5 and 1.4 GHz  
at
the resolution of 327 MHz observations may be contaminated by the
extended HII region.  In order to remedy this situation, we used
higher resolution 5 and 1.4 GHz images of the Arches cluster
with a resolution of 3.1$''\times3.8''$ and 3.2$''\times3.5''$ in order
to obtain a more accurate measure of radio continuum flux from the
Arches cluster.  The corresponding flux densities at 5 and
1.4 GHz are 11$\pm0.14$ and 15$\pm7$ mJy, respectively.  These estimates
assume that the emission from the background HII region is resolved, and 
has been removed  by background subtraction. The fluxes at
different frequencies are listed in Table 1 and are shown
in Figure 3. The flux values tend to be lower near 5GHz, indicating 
a transition region between thermal and nonthermal
components.

The high resolution continuum images resolve the Arches
cluster at 327, 1.4 and 5 and 8.5 GHz and 
deconvolved sizes  as listed in Table 1.  Considering the 
strong background emission at high frequencies and strong scattering
of the source at 327 MHz, the measured size of the source has large
uncertainties.  Nevertheless,  the measured size of
the radio emission appears to be consistent with a diffuse extended
component with a size approximately similar to the actual size of the
stellar cluster $\approx9''$.
The spectral index $\alpha$, where the flux density S$_{\nu} \propto
\nu^{\alpha} $, between 1.4 GHz and 327 MHz is determined to be
--1.2$\pm0.4$.  If we use low resolution 1.4 GHz data with
identical {\it uv} coverage to that of 327 MHz data, the flux density
is measured to be 88$\pm26$ mJy and the spectral index is estimated to
be $\alpha\approx-0.1\pm0.2$.  The low-resolution 1.4 GHz 
continuum image of the Arches cluster is clearly contaminated by the
background HII emission.  Taking the mean of these values, $\alpha$ is
estimated to be $\approx-0.6\pm0.4$,  consistent with a
nonthermal spectrum.

{\bf {X-ray Spectrum:}} 
 The  emission from A1 was fit with an absorbed,
two temperature model using a distant background region (Yusef-Zadeh et
al.  2002).  The majority of the observed flux (1/2-2/3) comes from the
low temperature component (kT $\sim$ 1keV).  However, there is
significant contribution to the flux, particularly the high energy line
emission, due to a  high temperature thermal component (kT $\sim$ 6
keV).  If the spectrum was fit with a one-temperature/powerlaw model,
the flux is dominated by the low temperature component (kT $\sim$
1.5keV), and only about 1/6th of the observed flux can be attributed to
the powerlaw.  The quality of both fits is roughly the same with
$\chi_{\nu}^{2}=0.65$ (using the $\chi^2$ Gehrels statistic, 
which  is more accurate  when the number of counts in each
bin is small),  whereas a
one-temperature model gives a $\chi_{\nu}^{2}=0.69$.  Unfortunately, the
\emph{un}absorbed flux from the powerlaw component is very difficult to
constrain.  The value of the photon index is basically unrestricted due to
the
relatively small contribution of the powerlaw to the total observed
flux.  Thus, the unabsorbed flux of the powerlaw component is
unconstrained. 

\section {Discussion}

The collision
of stellar winds in a dense cluster environment produces 
 X-ray emission from  a wind
escaping from the outer boundary of the cluster ( Cant\'o et al.\
2000, Raga et al.\ 2001).  Shocking of the most powerful stellar winds
produces discrete X-ray sources and the overall cluster outflow
produces diffuse X-ray emission.  The detection of diffuse X-ray
emission from the Arches cluster supported this picture (Yusef-Zadeh
et al.\ 2002).  These measurements suggest that nonthermal emission
can be produced by the acceleration of thermal X-ray gas to
relativistic energies by diffusive shock acceleration (e.g.\ Bell
1978).  Nonthermal emission could result from the contributions of
shocked stellar winds in binary systems of the cluster and shocked
cluster flow.  Nonthermal emission from massive binary systems have
been detected in many systems (e.g. Chapman et al.\ 1999), and is 
understood theoretically (Eichler \& Usov 1993).  The X-ray
luminosity from the Arches cluster is
$\approx(0.5-3)\times10^{35}$ whereas the nonthermal  and thermal 
luminosities are $4\pi
D^2 \nu F_\nu \approx 2.6\times10^{30}, \approx2\times10^{31}$
ergs s$^{-1}$, respectively.  The
ratio of X-ray to radio
luminosity is $\approx10^3$, similar to that detected in binary WR
binary systems.

The existence of nonthermal particles within the core of a luminous
young stellar cluster suggests that the nonthermal
X-ray or $\gamma$-ray emission could result from upscattering of the
radiation field by nonthermal particles (Ozernoy, Genzel \& Usov
1997).  
In fact, the Arches
cluster, is displaced by $\approx200''$ from the nominal position of
the unidentified EGRET source 3EG J1746--2851 (Hartman et al.\ 1999),
located well within the 95\% error radius $0.13^0$.  Other possible 
associations  
are with the radio Arc and G0.13-0.13 (Pohl et
al.\ 1997; Yusef-Zadeh et al.\ 2002b) where inverse Compton 
scattering of FIR radiation by the energetic electrons of the 
nonthermal filaments  and  brehmsstrahlung by the high-energy
tail of
the high electron energy distribution are used to reproduce  the
$\gamma$-ray emission.  
Both these suggestions suffer from  an
unestablished interaction
picture of the nonthermal filaments and adjacent molecular clouds. 
However, the Arches cluster is 
the closest candidate source to the centroid   of 3EG
J1746--2851 and the $\gamma$-ray emission is expected to arise from 
the cluster itself.  
Figure 4
shows the distribution of
radio and $\gamma$-rays from the region near the Arches cluster.  The
cross shows the position of 
3EG J1746--2851.  This steady,  strong
$\gamma$-ray source has  photon index 1.7$\pm0.6$ and a flux
1.2$\times10^{-6}$ photons cm$^{-2}$ s$^{-1}$ with energies greater
than 100 MeV. The $E^{-1.7}$ photon spectrum could
be produced by inverse Compton scattering from a distribution of
relativistic electrons emitting synchrotron radiation with a spectral
index of $\alpha$=--0.7, consistent with the 
 nonthermal radio emission from the Arches
cluster.

To estimate the gamma-ray flux produced by inverse Compton scattering,
we must first adopt parameters for the relativistic electron
population and the stellar radiation field within the Arches cluster.
We adopt a nonthermal synchrotron flux of 91 mJy at 327 MHz, with
spectral index $\alpha=-0.7$.  The radiating electrons then have an
$E^{-2.4}$
energy spectrum, which we assume extends between 10 MeV and 10 GeV.
The stellar radiation field in the Arches cluster is
approximated by a diluted black body with temperature $T =
3\times10^4$\,K; the dilution factor $W$ can be estimated from the
total cluster luminosity $L_{\mathrm{cl}} =10^{7.8}\lsol$ (Figer et
al.  2002), and cluster radius $r_{\mathrm{cl}} = 0.2$\,pc, as
$W\approx L / (4\pi\sigma r_{\mathrm{cl}}^2 T^4) = 1.1\times10^{-9}$.
With these nominal values, the inverse Compton flux above 100 MeV is
%\begin{equation}
   $ F(E>100 \u MeV ) = 0.87\times 10^{-6}
    \left( \frac{B\sin\beta }{20\mu\textrm{G}} \right)^{-1.7}
    \u ph \ut cm -2 \ut s -1 $
%\label{eq:F}
%\end{equation}
where $\beta$ is the pitch angle of the relativistic electrons. The
$\gamma$-ray flux decreases for stronger magnetic fields because fewer 
relativistic electrons are required to match the observed synchrotron
flux.  We conclude that if the EGRET source is a result of inverse
Compton scattering from the synchrotron-emitting electrons, the
magnetic field in the emitting regions is of order 20 microgauss.
This conclusion is insensitive to source geometry or filling factors
because the synchrotron and gamma-ray emitting regions are identical
and source geometry factors therefore scale out of equation (1).

Inverse Compton scattering may 
 contribute to the
X-ray emission from the center of the cluster, as 10 MeV
electrons scatter photons with $h\nu/k \approx 30\,000$ K
into the X-ray band.  For the numbers adopted above, the
band-integrated X-ray flux is $\approx (1\u keV )F(1\u keV )/0.3 =
1.0\ee -11 \u erg \ut cm -2 \ut s -1 $, roughly 1/4 of that from
X-ray component A1 (Yusef-Zadeh et al.\ 2002).
This is consistent with the one-temperature/power law fit to the spectrum
of A1 showing
that $\sim$1/6  of the observed flux can be attributed to the power law.

The ratio of magnetic field and relativistic electron energy densities
does depend on source geometry.  For a spherical source of radius 0.2
pc (5'' at 8.5 kpc), the inferred field strength is sub-equipartition:
the electron energy density is roughly a thousand times the energy
density in the field - the equipartition field (neglecting ions) would
be about 140 $\mu$Gauss.  Introducing a filling factor would increase
the equipartition field further.  The relativistic particle pressure,
while dominating the magnetic field, is several hundred times smaller
than the ram and thermal pressure associated with colliding winds from
the most massive stars in the cluster.  It is therefore plausible that
these particles are accelerated in shocks formed by colliding stellar
winds and are then advected outwards in the resultant cluster wind.

In summary, we have reported the first evidence of diffuse nonthermal
radio continuum emission from the Arches cluster, one of the youngest
and densest 
stellar clusters in the Galaxy.
The origin of $\gamma$-ray emission from 3EG J1746-2851 is accounted for
from
upscattering of intense radiation field in the cluster by 
relativistic particles generated from colliding winds. Future,
high-resolution observations using the VLA, Chandra, XMM and Integral
should be
able to examine the spectrum and the spatial distribution of
nonthermal emission from this cluster over a wide range of frequencies 
and determine the true spectrum of the Arches cluster.

%\special{landscape}
\ptlandscape

\makeatletter
\def\jnl@aj{AJ}
\ifx\revtex@jnl\jnl@aj\let\tablebreak=\nl\fi
\makeatother

\begin{deluxetable}{lllcccccccc}
\setcounter{table}{0}
\tablewidth{0pt}
\tablecaption{Gaussian Fits to the Arches Cluster}
\tablehead{
\colhead{Frequency} &
\colhead{Flux Density} &
\colhead{Background} &
\colhead{Deconvolved Size} &
\colhead{Resolution} & \cr
%\multicolumn{3}{c}{Size} & \cr
%\cline{8-10}
\colhead{} &
\colhead{(mJy)} &
\colhead{(mJy beam$^{-1}$)} &
\colhead{($''\times''$)} &
\colhead{($''\times''$)}}
%\colhead{PA ($^\circ$)} }
%\colhead{axial ratio}}
\startdata
330 MHz & 91$\pm$4.8 & 4.9 & 7.1$\times$13.5 &
7$\times$12&\nl
1.4 GHz & 15$\pm$7& -0.2 & 4.8$\times$6.2 &
3.2$\times$3.5&\nl
4.9 GHz & 11$\pm$0.1& 1.7 & 4.8$\times$6.6 &
3.1$\times$3.8&\nl
8.5 GHz & 14.8$\pm$4.8 & 8.8 & 5.5$\times$8.7 &
6.6$\times$7.8&\nl

\enddata
\end{deluxetable}

\vfill\eject
\begin{figure}
\plottwo{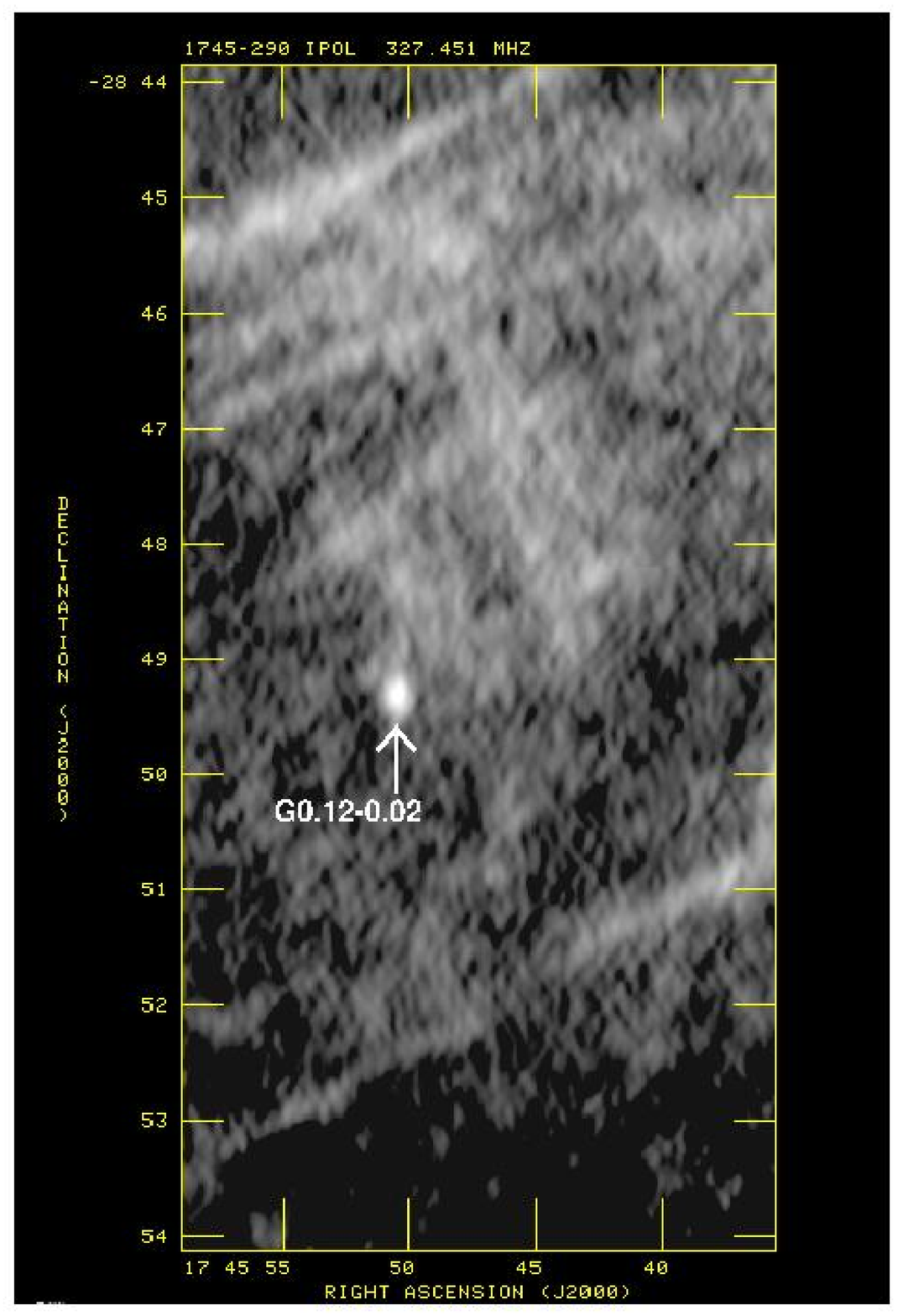}{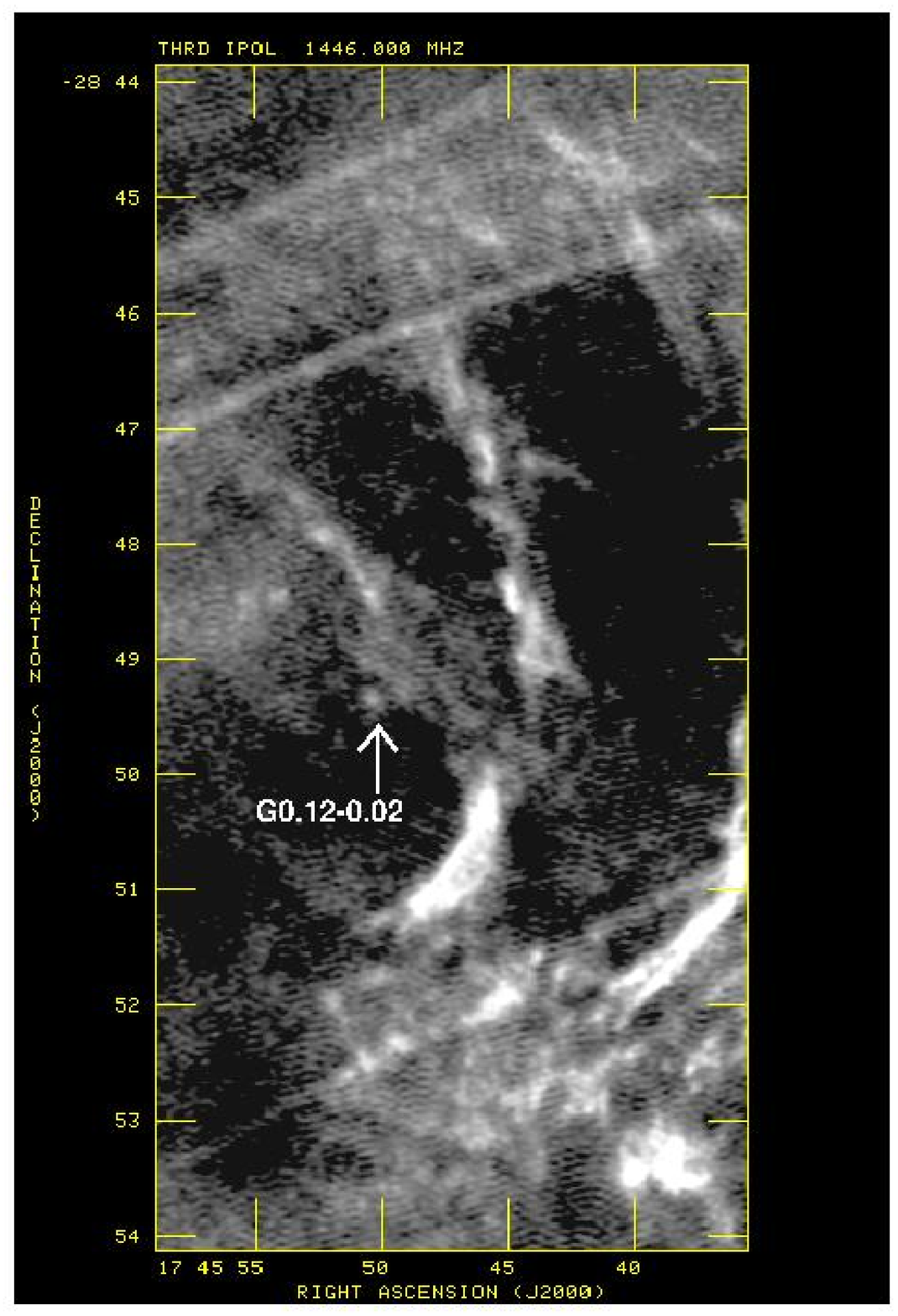}
\caption{Grayscale distribution of  327 MHz  and 1.4 GHz continuum
emission
from an identical region  are displayed in the left and right panels,
respectively. The peak emission at 327 MHz  coincides with the
Arches cluster.}
\end{figure}

\begin{figure}
\plotone{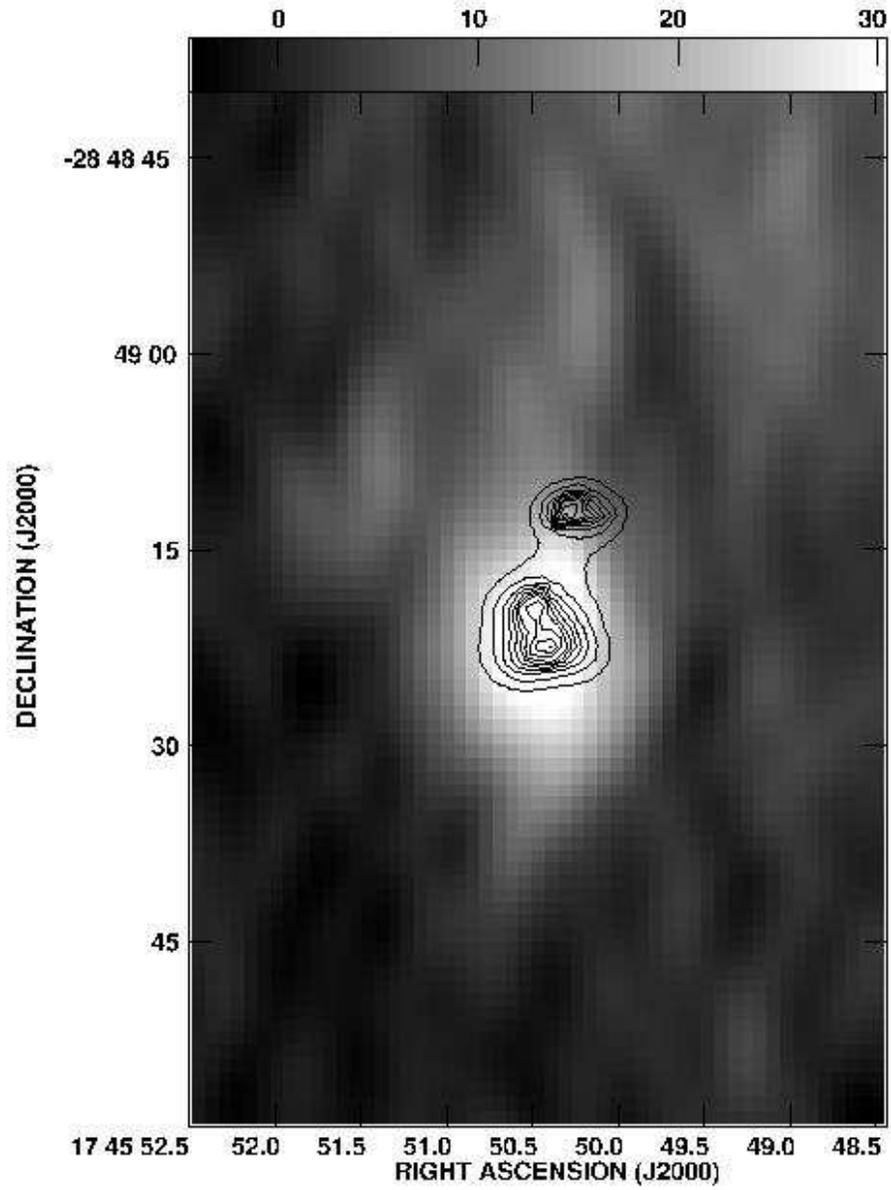}
\caption{Contours of X-ray emission based on Chandra observations
are superimposed on a grayscale distribution at 327 MHz.}
\end{figure}

\begin{figure}
\plotone{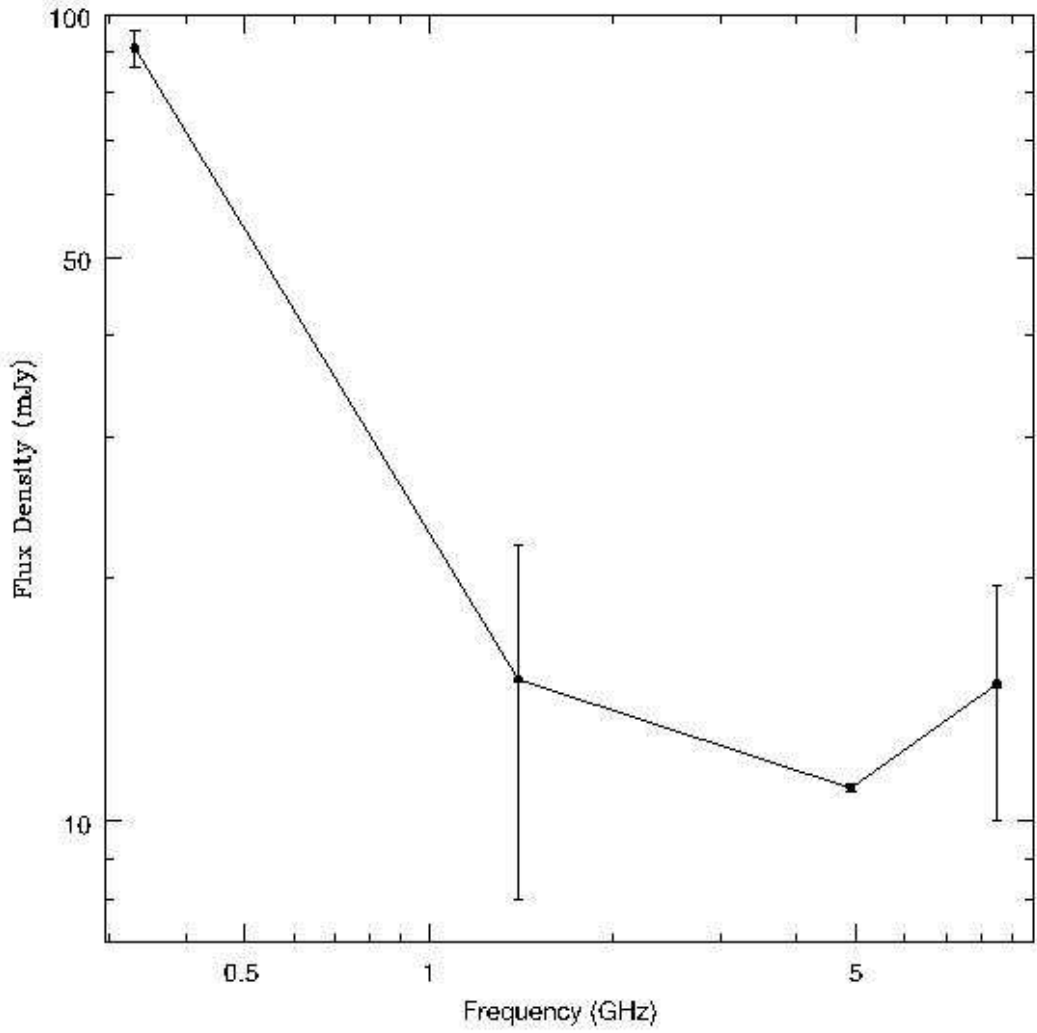}
\caption{A  radio spectrum of the Arches cluster between 327 MHz and 8 GHz}
\end{figure}

\begin{figure}
\plotone{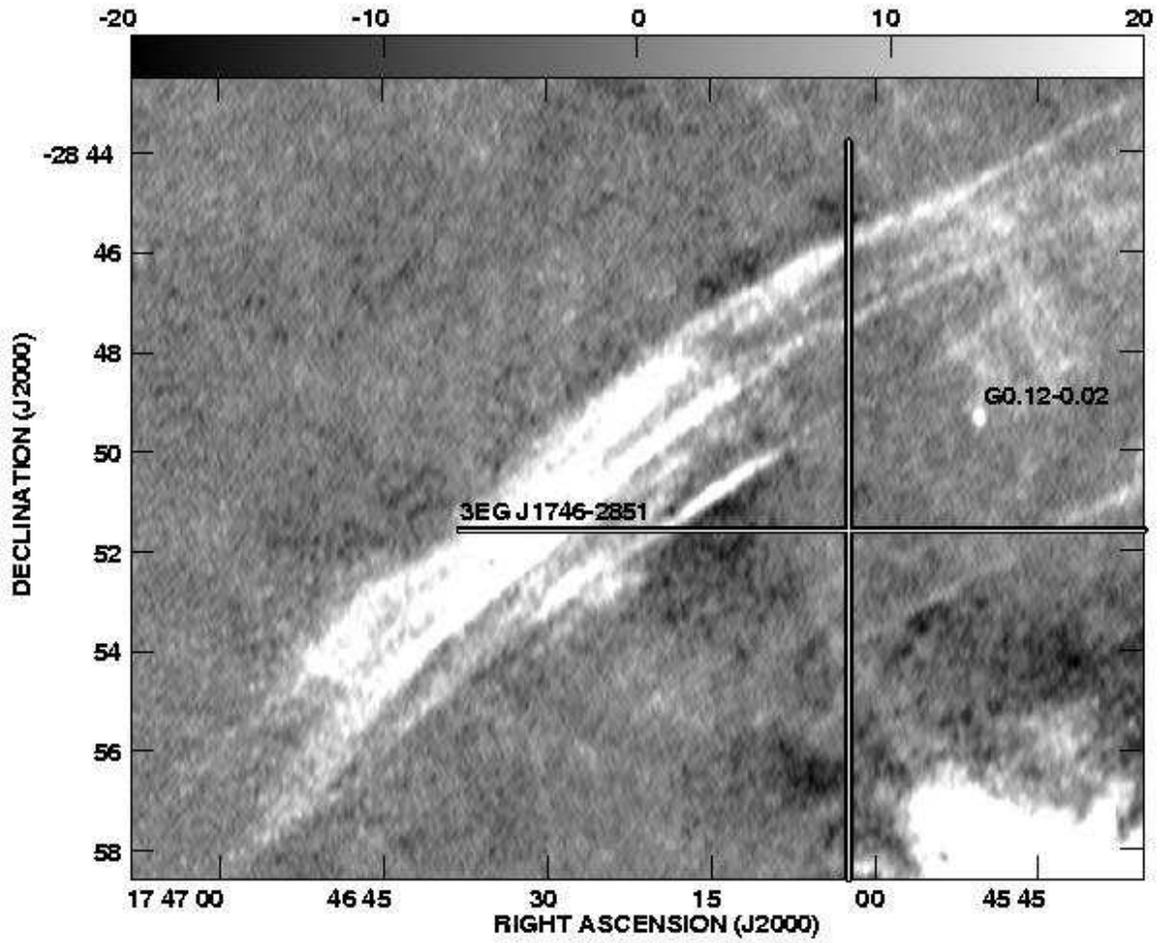}
\caption{
A grayscale 327 MHz image of the Arches cluster (G0.12-0.02), the
filaments of the radio Arc and the 
 Sgr A complex  to the southwest corner.  
3EG J1746-2851 is  drawn as  a cross with its  7.8$'$
radius of the 95\% error
circle. 
}
\end{figure}

\end{document}